\documentclass[aps,prl,twocolumn,reprint,showpacs,preprintnumbers,%
amsmath,amssymb,floatfix]{revtex4-1}
\usepackage{epsfig}
\usepackage{epsf}
\usepackage{amsmath}
\usepackage{amsfonts}
\usepackage{amssymb}
\usepackage{bm}
\usepackage{bbm}
\usepackage{nicefrac}
\usepackage{color}
\usepackage{pifont}
\usepackage{graphicx}
\usepackage{epsfig}
\usepackage{latexsym}

\def\d{\mathrm{d}}
\def\e{{\textrm e}}
\def\perm{\xi}
\newcommand{\paper}{Letter}
\newcommand{\colol}{(color online). }
\DeclareMathOperator{\tr}{Tr}
\def\vecb{\bm}

\begin{document}

\title{Triple Compton effect:\\
A photon splitting into three upon collision with a free electron}
\date{\today}
\author{Erik L\"otstedt}
\email{lotstedt@chem.s.u-tokyo.ac.jp}
\altaffiliation[On leave of absence from:]{Department of Chemistry, School of Science,%
The University of Tokyo, 7-3-1 Hongo, Bunkyo-ku, Tokyo 113-0033, Japan}
\author{Ulrich D. Jentschura}
\affiliation{Department of Physics, 
Missouri University of Science and Technology, 
Rolla, Missouri 65409-0640, USA}

\begin{abstract}
The process in which a photon splits into three after the collision with a free
electron (triple Compton effect) is the most basic process for the generation 
of a high-energy multi-particle entangled state composed out of elementary quanta.
The cross section of the process is evaluated in two experimentally realizable
situations, one employing gamma photons and stationary electrons, and the other
using keV photons and GeV electrons of an x-ray free electron laser. For the
first case, our calculation
is in agreement with the only  available measurement of the differential
cross section for the process under study. Our estimates indicate
that the process should be readily  measurable
also in the second case. 
We quantify the polarization
entanglement in the final state by a recently proposed 
multi-particle entanglement measure. 
\end{abstract}


\maketitle

%
%

{\it Introduction.}---The triple Compton effect is a fundamental process of
light-matter interaction, in which a photon splits into three upon 
collision with an electron, 
\begin{equation}
\label{tripleComptoneffect}
\e^-+\gamma \longrightarrow \e^-+\gamma +\gamma +\gamma.  
\end{equation} 
The cross section of this 
quantum electrodynamical (QED) process is of fourth order in the fine-structure
constant $\alpha\approx 1/137.036$.  Currently,
there is no general theoretical
treatment of the triple Compton effect in the literature. In principle, there
is no  limit of the number of photons that can be coherently emitted when an
electron interacts with a photon, but the only processes which have been
discussed so far are the usual, second-order (in $\alpha$) single Compton 
effect~\cite{KlNi1929} and the third-order double Compton effect
\cite{MaSk1952}. Only the single Compton effect   has a classical limit.
To the contrary, both the double and the triple Compton effects are
intrinsically quantum processes which cannot be described by classical
electrodynamics.  With strong lasers becoming available, the nonlinear
generalization of Compton scattering in which several laser photons are
absorbed have been vigorously discussed; see
\cite{HaHeIl2009,MaDPKe2010} for recent
investigations of the nonlinear Compton effect, and
\cite{LoJe2009prl,LoJe2009pra,SeKe2012} for a discussion on the nonlinear
double Compton effect, where two photons are coherently emitted.

The three photons emitted in the process \eqref{tripleComptoneffect} originate
from the same initial photon, are emitted at the same time, and are therefore
quantum mechanically entangled. The creation of an entangled state of three
photons is an important goal in quantum information. The conventional way of
generating entangled triple-photon states is by employing nonlinear crystals
\cite{PaEtAl2000,HuEtAl2000,AnGeKr2011}, but one
can imagine other sources, not yet experimentally realized,  such as
electron-positron annihilation \cite{Gu1955,AbAdYo2011}. In this
\paper, we propose the triple Compton effect as an alternative source of
entangled photon triplets, which by suitable optimization could be competitive
both in terms of production rate and the degree of entanglement. Especially
interesting is the possibility of creating correlated photons with high energy
in the GeV range, via triple Compton backscattering on a relativistic electron
beam. We will show that a advantageous setup for such an experiment is  an
x-ray free electron laser (XFEL), providing a high-flux, high-energy photon
beam together with a GeV electron beam.

The only previous experimental study of the triple Compton effect that we are
aware of is Ref.~\cite{MGBr1968}, where the differential cross section 
was measured for a well-defined interval of the solid angle, 
as defined by the detectors which were arranged in a symmetric 
angular configuration.
On the 
theoretical side,  the only preceding investigation is \cite{MaMaDh1959},
where the total cross section in the  limit of ultra-relativistic initial
photon energy  was studied.  In our treatment of the problem, we can evaluate
the differential cross section for arbitrary values of the directions, energies
and polarizations of the emitted photons. On the other hand, the double Compton
effect is rather well studied, both 
experimentally~\cite{MGBrKn1966,SaEtAl2000,SaSiSa2008,SaSiSa2011} and 
theoretically~\cite{MaSk1952,RaWa1971,Go1984}. In addition, several other
processes have been studied where two photons are produced in the final state,
such as double bremsstrahlung~\cite{BaEtAl1981,KrMaMaSt2002,KoSo2006},
bound state decay~\cite{RaSuFr2008epjd,JeSu2008}, and laser-induced
photon splitting~\cite{DPMiKe2007}, but comparatively little is known about
triple photon production.

%
%
{\it Theoretical formulation.}---Unless stated otherwise, natural units with
$\hbar=c=1$ are used, and $m$ denotes the electron mass.  Four-vector
products are denoted with a dot 
(i.e., we have
$a\cdot b \equiv a^\mu \, b_\mu = 
a^0 \, b^0-\vecb{a}\cdot \vecb{b}$ for two four-vectors $a$ and $b$). 
The contraction with the Dirac
matrices $\gamma^\mu$ is written with a hat, $\hat{a}=a^0\gamma^0-\vecb{a}\cdot
\vecb{\gamma}$.
 
\begin{figure}
\includegraphics[width=0.93\linewidth]{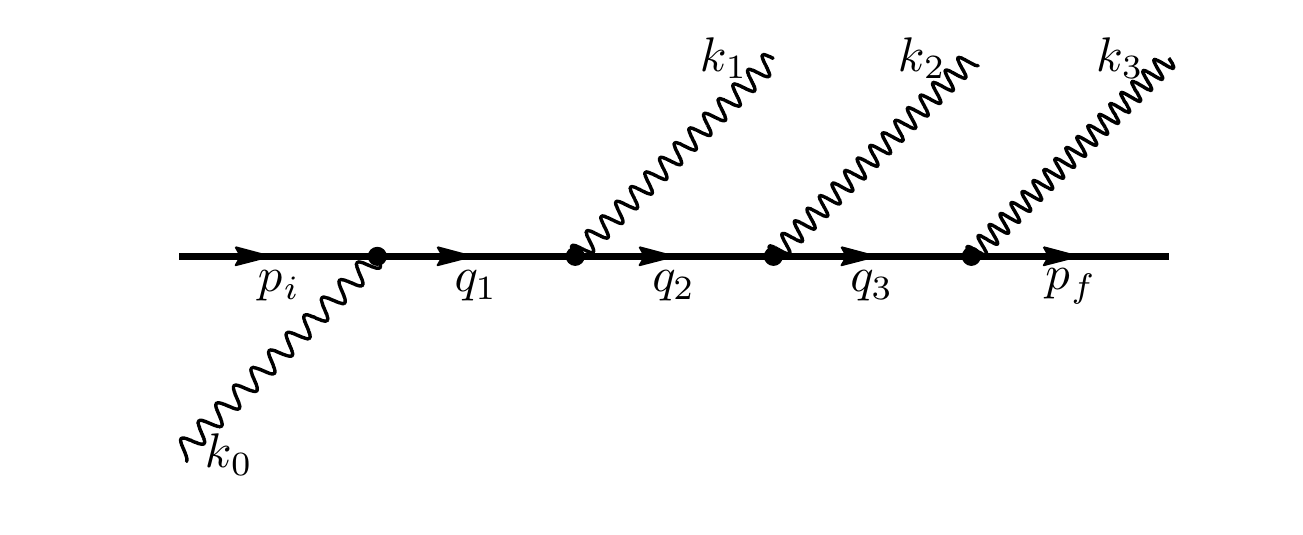}
\caption{\label{Feynmandiag}One of the $4!=24$ contributing Feynman
diagrams of the triple Compton effect. To obtain the total amplitude, we have
to add the diagrams corresponding to the permutations of the insertion order of
the photons $k_0$, $k_1$, $k_2$, and $k_3$.}
\end{figure}

We label the four-momenta of the incoming electron and photon with
$p_i=(E_i,\vecb{p}_i)$ and $k_0=\omega_0 \, n_0 = 
\omega_0 \, (1,\vecb{n}_0)$,
respectively, and the four-momenta of the outgoing electron and photons with
$p_f=(E_f,\vecb{p}_f)$, and $k_j=\omega_j n_j=\omega_j(1,\vecb{n}_j)$,
$j=1,2,3$, respectively. When explicitly evaluating the cross section, we take
the vectors $\vecb{k}_j=\omega_j\vecb{n}_j$ in the lab frame, in spherical coordinates with the
polar axis directed along 
 $\vecb{k}_0$, i.e., $\vecb{k}_0=\omega_0(0,0,1)$ and
$\vecb{k}_j=\omega_j(\sin\theta_j\cos\phi_j,\sin\theta_j\sin\phi_j,\cos\theta_j)$.
The amplitude $M$ for the triple Compton effect is formed from a coherent sum
of $4!=24$ Feynman diagrams, one of which is shown in Fig.~\ref{Feynmandiag}. 

%
%

We have  
\begin{align}
\label{defN}
M=m^3
\sum_{\perm} 
&
u_{r_f}^\dagger(p_f)\gamma^0 
\hat{\epsilon}_{\perm(3)}\frac{\hat{q}_3 (\perm)+m}{q_3^2(\perm)-m^2}
\hat{\epsilon}_{\perm(2)}\frac{\hat{q}_2(\perm)+m}{q_2^2(\perm)-m^2}
\nonumber
\\&
\times
\hat{\epsilon}_{\perm(1)}\frac{\hat{q}_1^2(\perm)+m}{q_1^2(\perm)-m^2}
\hat{\epsilon}_{\perm(0)}u_{r_i}(p_i) \,,
\end{align}
where the sum runs over all the 24 permutations $\perm$ of $(0,1,2,3)$,
and $r_f$ is the final electron spin.  The
momenta $q_n$ entering the propagators are defined as 
$
q_n(\perm)=p_i + \sum_{j=0}^{n-1}(-1)^{\delta_{0\perm(j)}+1}\,k_{\perm(j)},
$
where $\delta_{ij}$ is Kronecker's delta function. 
(One adds to $p_i$ the four-momentum of the incoming photon
$k_0$ and subtracts the four-momenta of the outgoing photons, 
according to the relevant permutation $\perm$.)
In the expression for 
the amplitude \eqref{defN}, we have introduced the 
positive-energy bispinor $u_r(p)$, 
with the spin index $r=1,2$, which we use in the 
conventions of Chap.~2 of~\cite{ItZu1980} (the normalization is
$\overline u_r(p) \, u_r(p) = 
 u^\dagger_r(p) \, \gamma^0 \, u_r(p) = 1$).
The polarization four-vectors 
$\epsilon_j=(0,\vecb{\epsilon}_j)$ satisfy
$\epsilon_j\cdot k_j=0$ for $j=0,1,2,3$. 
As an explicit set of basis vectors for~$\vecb{\epsilon}_j$,
we take~$\vecb{\epsilon}_j^1=
(\cos\theta_j\cos\phi_j,\cos\theta_j\sin\phi_j,-\sin\theta_j)$ and 
$\vecb{\epsilon}_j^2=
(-\sin\phi_j,\cos\phi_j,0)$.
The differential cross section 
follows from the usual rules of QED \cite{JaRo1980} as 
\begin{align}
\label{diffcrosssectionfinal} 
\frac{\d\sigma}{\d\omega_1 \d\omega_2 \d\Omega_1 \d\Omega_2 \d\Omega_3}
\equiv \sigma_5={}
&
\frac{\alpha^4}{(2\pi)^4} 
\frac{1}{m^4}
\frac{\omega_1 \, \omega_2 \, \omega_3}{ E_f \, p_i \cdot k_0} 
\frac{|M|^2}{|K|} 
\nonumber
\\
&
\times
\Theta(\omega_3)\Theta(E_f-m),
\end{align}
where $\Theta(\cdot)$ is the 
step function, $\d\Omega_j=\sin\theta_j\d\theta_j \d\phi_j$ 
is the differential solid angle of 
photon $j$, and we have introduced $\sigma_5$ as an abbreviation of the five-fold 
differential cross section. In Eq.~\eqref{diffcrosssectionfinal}, 
the final four-momentum of the electron is fixed by energy-momentum 
conservation as 
$p_f=p_i+k_0-k_1-k_2-k_3$,  
and the energy of photon three is 
\begin{equation}
\label{omega3}
\omega_3=\frac{p_i\cdot( k_0- k_1- k_2)-k_0\cdot ( k_1+ k_2)+k_1\cdot k_2}%
{n_3\cdot (p_i+k_0-k_1-k_2)}.
\end{equation}
The factor $K$ in \eqref{diffcrosssectionfinal}, arising from the final delta
function integration over $\omega_3$ reads $K = 1 + \frac{ \vecb{n}_3\cdot 
(\vecb{k}_1+\vecb{k}_2-\vecb{k}_0-\vecb{p}_i)+\omega_3}{E_f}$.

The differential cross section \eqref{diffcrosssectionfinal} depends on eight
continuous variables, two angles $\theta_j$, $\phi_j$ for each emitted photon
and the energies $\omega_1$, $\omega_2$ of two of the emitted photons. The
energy $\omega_3$ of the third photon is restricted by the kinematical
constraints and is calculated by Eq.~\eqref{omega3}. In addition, 
we have six more discrete variables which can take any of two 
values, namely, the
polarization vectors $\vecb{\epsilon}_j$ of the photons and the spins $r_{i,f}$
of the electrons.

A couple of general remarks about the expression \eqref{diffcrosssectionfinal}
are the following.  Similarly to the double Compton effect
\cite{JaRo1980}, the cross section vanishes when all three photons are
emitted parallel to the incoming photon. 
In the rest frame of the electron, 
this implies 
$\omega_1+\omega_2+\omega_3=\omega_0$  
according to Eq.~\eqref{omega3}, 
 and therefore corresponds to the incoming
photon splitting into three photons without any interaction, which is not allowed.
Whenever either of $\omega_1$, $\omega_2$ or $\omega_3$ goes to zero,
$\sigma_5$ diverges as $1/\omega_{1,2,3}$,
while the radiated energy remains finite in the infrared. 
This is the well-known infrared
catastrophe of QED, which can be cured by adding radiative 
corrections~\cite{BrFe1952}. In the current case, the 
divergences at $\omega_n\to 0$ would
cancel against radiative corrections to the single and double Compton effect. The
evaluation of such corrections is beyond the scope of this \paper. We will
calculate the differential cross section sufficiently far from the infrared
divergences, corresponding to a specific experimental detector threshold.
Our calculations therefore have a relative accuracy of the order 
of the fine-structure constant.

%
%
{\it Evaluation of the differential cross section.}---The  
evaluation of $\sigma_5$ is performed numerically, 
by employing an explicit representation of $\gamma^\mu$ and $u_r(p)$.
 This approach is advantageous \cite{ScLoJeKe2007pra} if one is 
interested in polarization-resolved cross sections, in which case the analytic 
evaluation of $\sigma_5$ does not simplify.
We 
calculate the
differential cross section \eqref{diffcrosssectionfinal} for two
different experimental setups.  The first is the same as considered 
in Ref.~\cite{MGBr1968}, were a measurement of the
triple Compton cross section is described. In this setup,
a gamma photon of energy $\omega_0=0.662$~MeV impinges on an electron at rest,
$E_i=m$, and photons are detected in coincidence at
$\theta_1=\theta_2=\theta_3=\pi/2$, and $\phi_1=2\pi/3$, $\phi_2=4\pi/3$,
$\phi_3=0$.  The photon energy threshold was $\omega_n\geq \varepsilon=13$~keV.
What was actually measured was the differential cross section
averaged over the solid angles subtended by the detectors, 
\begin{equation}\label{average_sigma}
\langle \sigma \rangle=
\frac{1}{\Omega^3}\frac{1}{4}
\sum_{\textrm{spin},\,\textrm{pol.}}
\int_\Omega \d\Omega'_1 \d\Omega'_2 \d\Omega'_3 
\int \d\omega_1 \d\omega_2
\sigma_5,
\end{equation}
where the solid angle is $\Omega=0.378$\,sr, and the energy integration is over 
$\omega_{1,2,3} > \varepsilon$.  Performing the integration in
\eqref{average_sigma} over the angular intervals
$\Delta\phi_j=\phi_j\pm\sqrt{\Omega}/2$, $\Delta\theta_j=\theta_j\pm
\arcsin(\sqrt{\Omega}/2)$ so that
$\int_{\Delta\theta_j}\d\cos\theta'_j\int_{\Delta\phi_j}\d\phi'_j=\Omega$, and
over photon energies greater than $\varepsilon$, we obtain 
\begin{equation}
\label{theor}
\langle \sigma \rangle=4.1\times 10^{-9}\, {\rm b}/{\rm sr}^3 \,.
\end{equation}
This value should be compared to the measured value of 
\begin{equation}
\label{expmt}
\langle \sigma \rangle^{\textrm{exp}}=(8.1\pm2.4)\times 10^{-9}\,
{\rm b}/{\rm sr}^3 \,,
\end{equation}
which includes an experimental uncertainty of 30\%~\cite{MGBr1968}. 
Although the calculated value lies outside the estimated
experimental error bar (by $1.6$
standard deviations), we believe that our calculation rather can  be regarded as a
confirmation of the measurement in Ref.~\cite{MGBr1968}, given the 
utmost difficult nature of the experiment.
We have extended the analysis relevant to the experiment~\cite{MGBr1968}
to the energy interval $\omega_0 \in (10^{-2}, 10^2 )$~MeV;
details will be presented elsewhere.

The second example we consider is a triple Compton backscattering scheme: 
The head-on collision of a relativistic electron, 
at an incident energy of $E_i=5$\,GeV and $\vecb{p}_i=(0,0,-\sqrt{E_i^2-m^2})$, 
with an incoming photon of energy $\omega_0=1$~keV. Such parameters are realized 
in an XFEL, for example at the 
LCLS in Stanford~\cite{LCLS}, and would require the 
reflection of the x-ray beam to collide with the 
electron beam from the accelerator.
In this situation, the photons are 
backscattered and emitted in a narrow cone around the 
propagation direction of the electron, $\theta_j\sim\pi-E_i/m$. 
We estimate the total cross section as   
$\sigma_{\rm TC}^{\textrm{tot}}=2\times 10^{-5}$~b, compared 
to $\sigma_{\rm DC}^{\textrm{tot}}=1\times 10^{-3}$~b and 
$\sigma_{\rm SC}^{\textrm{tot}}=3\times 10^{-2}$~b, 
where for DC and TC, we have assumed that 
all photons with
energies larger than the assumed 
photon energy threshold $\varepsilon=50$\,MeV are detected. 
If we adopt pulse parameters available at the LCLS~\cite{LCLS}, i.e.,
$2\times 10^{13}$ photons per pulse, $10^9$ electrons per bunch, 
a transverse bunch size of $40$\,$\mu$m, perfect transverse overlap of 
the two pulses, and a repetition rate of 120\,Hz, we obtain
3 triple photon events per second.

%
%
\begin{figure}
\includegraphics[width=0.93\linewidth]{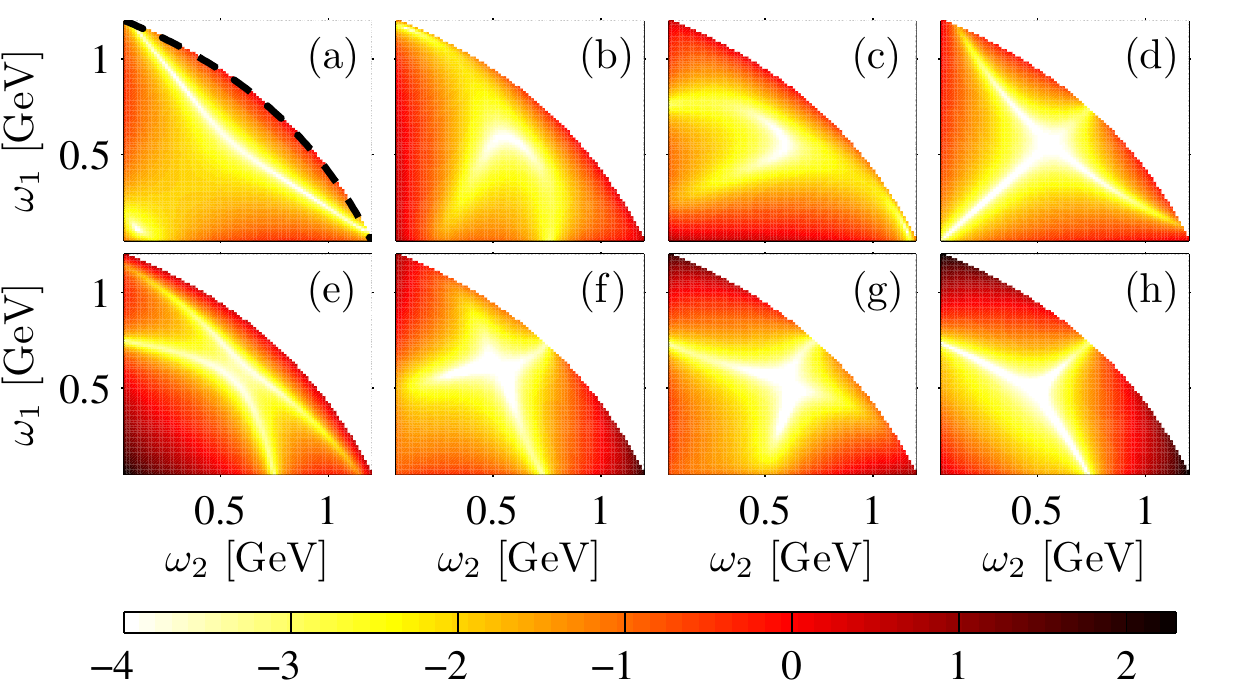}
\caption{\label{fig3}\colol The five-fold differential cross section 
at $\theta_j=\pi-1.5\times 10^{-3}$ and $\phi_j=2j\pi/3$, $j=1,2,3$ as 
a function of $\omega_1$ and $\omega_2$. The color indicates the value 
of the cross section $\log_{10}(\frac12 \,
\sum_{\textrm{spin}}\sigma_5/\textrm{b}\,\textrm{MeV}^{-2}\,\textrm{sr}^{-3})$
summed over the electron spins. We have $E_i=5$~GeV and 
$\omega_0=1$~keV, and the incoming photon 
beam is linearly polarized in the $x$-direction.
The different panels correspond  to different polarization of the final photons, 
we have $(\vecb{\epsilon}_1,\vecb{\epsilon}_2,\vecb{\epsilon}_3)=
(\vecb{\epsilon}_1^j,\vecb{\epsilon}_2^k,\vecb{\epsilon}_3^l)$,
with $jkl=$ (a) 111, (b) 211, (c) 121, (d) 112, (e) 221, (f) 212, 
(g) 122 and (h) 222. 
The black dashed line in panel (a) shows the curve along which the energy of 
photon three is at the assumed detector threshold, 
$\omega_3=\varepsilon=50$~MeV.}
\end{figure}

One example of the fully differential cross section in the XFEL setup is shown
in Fig.~\ref{fig3}. In this figure, we assume a configuration 
centered near the incoming electron axis, with $\theta_j=\pi-1.5\times
10^{-3}$  
and $\phi_j=2j\pi/3$, $j=1,2,3$. 
We evaluate $\sigma_5$ as a function
of $\omega_1$ and $\omega_2$. We sum over the electron spins, but assume a
linearly polarized XFEL beam, and also fix the polarization of the emitted
photons. We show $\sigma_5$ only for $\omega_{1,2,3}>\varepsilon$.
For photon energies smaller than $\varepsilon$, the differential cross section
is set to zero.  It becomes clear from Fig.~\ref{fig3} that high energy
photons in the GeV range can be achieved in this setup, which is interesting
since it could provide correlated photon triplet states, with quantifiable
entanglement as shown below, in an energy range which is far beyond what can be
produced with down-conversion of photons in a nonlinear crystal
\cite{PaEtAl2000,HuEtAl2000,AnGeKr2011}.  
The rate of 3 events/s obtained above is not small compared to the
experimental results in \cite{HuEtAl2000}, where photon triplets produced
from nonlinear down-conversion were detected at an event rate of 5 per hour.
The polarization of GeV photons can be measured using coherent electron-positron 
pair production in aligned crystals \cite{ApEtAl2005}.

%
%
{\it Entangled photon triplets.}---The three photons in the final state are 
simultaneously emitted and their polarizations are inevitably entangled due to 
the quantum nature of the process. In general, 
the mixed polarization state resulting after the interaction can be described by 
the density matrix $\varrho$, which has matrix elements 
\begin{equation}
\langle \lambda_1 \lambda_2 \lambda_3 | \varrho | 
\lambda'_1 \lambda'_2 \lambda'_3 \rangle =
N\sum_{r_i,r_f}  M(\lambda_1\lambda_2\lambda_3) \,
M^\ast( \lambda'_1 \lambda'_2 \lambda'_3),
\end{equation}
where we have written $M(\lambda_1\lambda_2\lambda_3)=
M(\vecb{\epsilon}_1=\vecb{\epsilon}_1^{\lambda_1},\vecb{\epsilon}_2=\vecb{\epsilon}_2^{\lambda_2},
\vecb{\epsilon}_3=\vecb{\epsilon}_3^{\lambda_3})$, $\lambda_j\in \{1,2\}$, and
the normalization constant $N$ is 
chosen so that $\tr(\varrho) =1$ (we sum over initial and final
electron spins $r_i$ and $r_f$).  It is still an
open problem and subject of active research 
how to quantify the entanglement of a given multi-particle
state $\varrho$
(see Refs.~\cite{ToGuSeUf2005,BaGiLiPi2011,%
JuMoGu2011}). In
the current study, we employ the entanglement measure put forward in
\cite{JuMoGu2011} in order to estimate the degree of
polarization entanglement present among the final three photons.  Briefly, an
entanglement witness $W$  is found by minimizing the trace $\tr (W\varrho)$
with $W$ such that for all subsets $s$, $W=P_s+Q_s^{T_s}$, where $T_s$ denotes the partial
transpose with respect to the subset $s$ 
\cite{LeKrCiHo2000}, and   the matrices $P_s$, $Q_s$, $1-P_s$ and
$1-Q_s$ should have positive eigenvalues. Then,
$
\tau(\varrho)=-\tr (W\varrho)
$
is a measure of the tripartite polarization entanglement present in $\varrho$.  In
particular, $\tau=0$ for states which are not genuinely tripartite entangled.
As an example, $\tau(\varrho_{\textrm{GHZ}})=1/2$ for the 
Greenberger--Horne--Zeilinger (GHZ) state
$\varrho_{\textrm{GHZ}}=|\textrm{GHZ}\rangle \langle \textrm{GHZ}|$, with
$|\textrm{GHZ}\rangle =(|111\rangle+|222\rangle)/\sqrt{2}$
\cite{GrHoShZe1990}, so that $\tau=1/2$ may be regarded
as maximum entanglement.  

%
%
\begin{figure}
\includegraphics[width=0.93\linewidth]{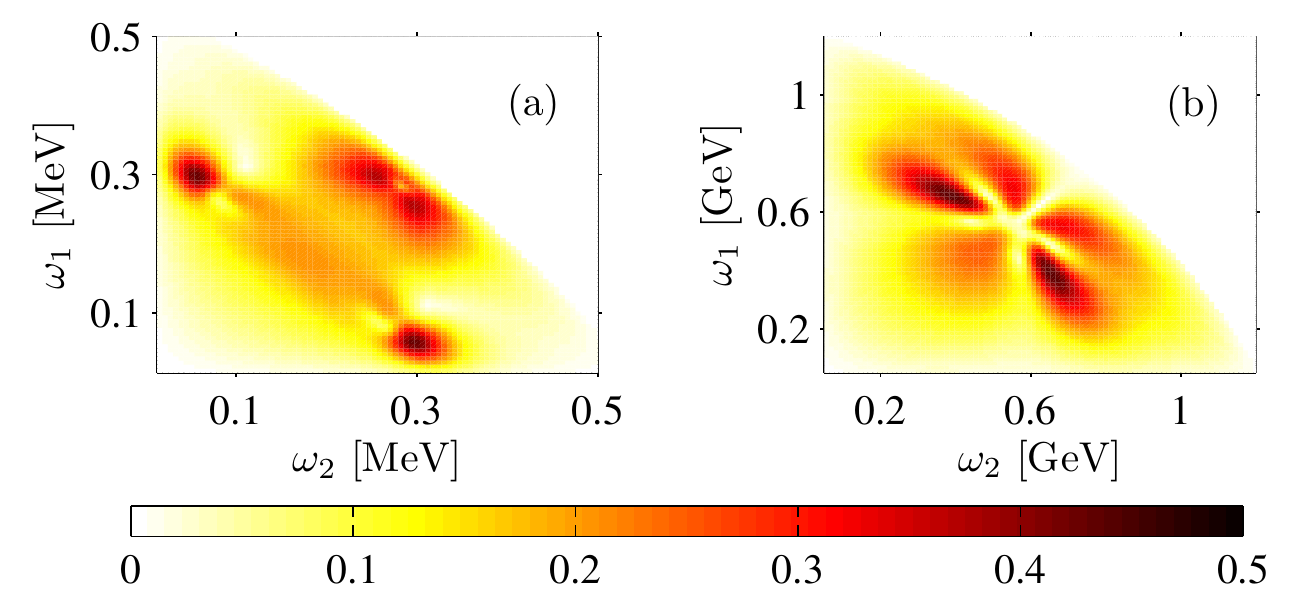}
\caption{\label{fig4}\colol 
Entanglement measure $\tau$ as a function 
of the photon energies $\omega_1$ and $\omega_2$, for 
(a) $\omega_0=0.662$~MeV, $E_i=m$, $\theta_{j}=\pi/2-1$, 
$\phi_j=2j\pi/3$, $j=1,2,3$, $\varepsilon=13$~keV, and (b) 
$\omega_0=1$~keV, $E_i=5$~GeV, $\theta_{j}=\pi-1.5\times 10^{-3}$, 
$\phi_j=2j\pi/3$, $j=1,2,3$, $\varepsilon=50$\,MeV. 
In both cases spin was summed over, and 
the initial photon beam was assumed to be 
linearly polarized in the $x$ direction. 
The differential cross section, and thus also $\tau$, was set to zero  
for photon energies below the detector threshold $\varepsilon$.}
\end{figure}

In Fig.~\ref{fig4}, we show the value of $\tau$, evaluated as a function of
$\omega_1$ and $\omega_2$ at (a) $\omega_0=0.662$~MeV, $E_i=m$,
$\theta_{1,2,3}=\pi/2-1$, and $\phi_j=2j\pi/3$, and (b) $\omega_0=1$~keV, $E_i=5$~GeV,
$\theta_{1,2,3}=\pi-1.5\times 10^{-3}$, and $\phi_j=2j\pi/3$ (the same
situation as in Fig.~\ref{fig3}).  The numerical evaluation of $\tau$ was
performed with {\sc pptmixer}~\cite{JuMoGu2011}, available at
\cite{pptmixer}.  In both cases a non-zero value of $\tau$ for almost all
values of $\omega_1$, $\omega_2$ where $\sigma_5>0$ shows that the three
photons are indeed entangled.  Almost maximum entanglement $\tau=1/2$ is
achieved at certain places in the $\omega_1\omega_2$ plane.

%
%
{\it Conclusions.}---We have presented a theoretical study of the triple
Compton effect, where a photon is split into three after colliding with an
electron. Our formulation of the problem enables us to evaluate the
differential cross section at arbitrary angles and energies of both initial and
final particles. We verify the 44-year old experimental 
result reported in Ref.~\cite{MGBr1968}.
Theory and experiment are not in perfect agreement, but the 
discrepancy of roughly $1.6$ standard deviations is not significant
[see Eqs.~\eqref{theor} and~\eqref{expmt}]. 
Additional measurements are needed to clarify this issue.
A straightforward generalization of the formalism to 
a Compton backscattering geometry then leads to theoretical predictions
for the triple Compton effect for typical parameters
at modern XFEL facilities, as shown in Fig.~\ref{fig3}.
Finally, while it is intuitively rather clear what a 
two-particle entangled quantum state is,
the quantification of three-particle entanglement is a 
much more subtle problem. The entanglement measure $\tau$
has been proposed in 
Ref.~\cite{JuMoGu2011},
and corresponding theoretical predictions in Fig.~\ref{fig4}
show that the generation of a multi-particle entangled 
photon state in the MeV or GeV regime is possible 
based on the triple-Compton effect as a basic three-quanta emission 
process. The triplet photon final state is polarization
entangled. 

%
%
\acknowledgments
We acknowledge support from the National Science Foundation 
and from the National Institute of Standards and Technology 
(NIST precision measurement grant).

\end{document}